\magnification=1200
\baselineskip=18truept
\input epsf

\def\preprint{Y}
\def\draftversion{N}
\def\cap{\hsize=4.6in}

\if \draftversion Y


\fi

\def\figure#1#2#3{\if \preprint Y \vskip .5cm \midinsert \epsfxsize=#3truein
\centerline{\epsffile{figure_#1_eps}} \halign{##\hfill\quad
&\vtop{\parindent=0pt \hsize=4.6in \strut## \strut}\cr {\bf Figure
#1.}&#2 \cr} \endinsert \fi}

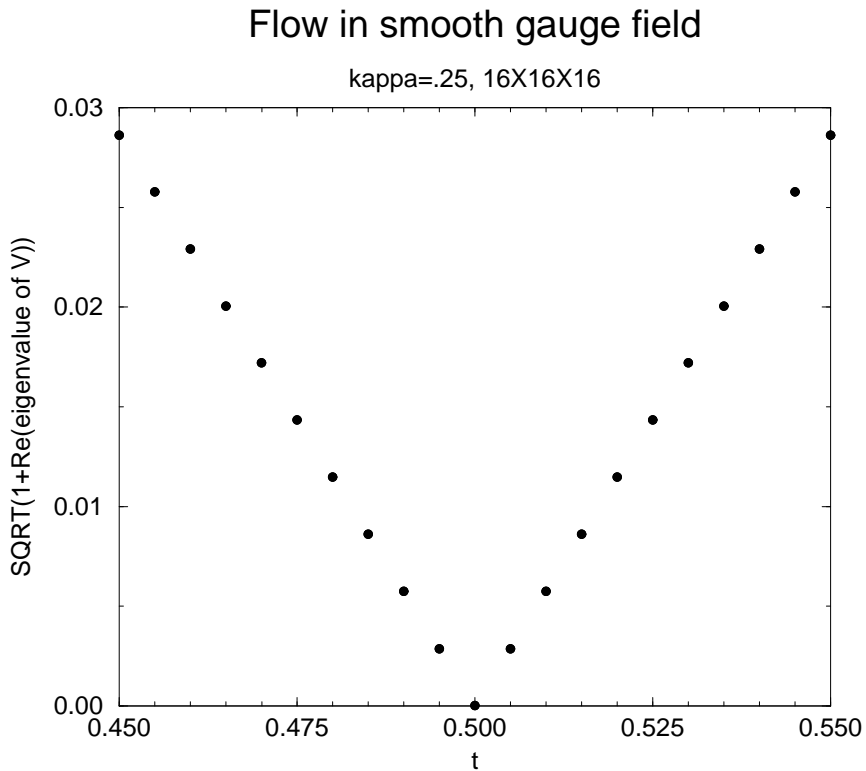
\def\captionone{\cap The level crossing.}
\def\plbfirst{1}
\def\kaplan{2}
\def\frolovslavnov{3}
\def\npblong{4}
\def\twod{5}
\def\exactlymassless{6}
\def\GW{7}
\def\hasen{8}
\def\luscher{9}
\def\appel{10}
\def\casherneub{11}
\def\twchiu{12}
\def\higham{13}
\def\multiplemass{14}
\def\odddim{15}
\def\realsu2{16}
\def\su2anom{17}
\def\ritz{18}
\def\berry{19}
\def\lepage{20}
\def\trunca{21}
\def\truncb{22}
\def\rebbi{23}

\line{\hfill RU--98--28}
\vskip 2cm
\centerline {\bf A practical implementation of the Overlap-Dirac
operator.}
\vskip 1cm
\centerline{Herbert Neuberger}
\vskip .25cm
\centerline{\tt neuberg@physics.rutgers.edu}
\vskip 1.5cm
\centerline{\it Department of Physics and Astronomy}
\centerline{\it Rutgers University}
\centerline{\it Piscataway, NJ 08855--0849}
\vskip 2cm
\centerline{\bf Abstract}
\vskip .5cm
A practical implementation of the Overlap-Dirac operator
${{1+\gamma_5\epsilon(H)}\over 2}$ is presented. The implementation
exploits the sparseness of $H$ and does not require full storage.
A simple application to parity invariant
three dimensional $SU(2)$ gauge
theory is carried out to establish that zero modes related 
to topology are exactly reproduced on the lattice.

\vfill
\eject
The preservation of global chiral symmetries on the lattice has
been a long standing problem. During the last five years a solution
has been found but in the initial stages of the finding practical
applicability seemed limited to two dimensions (2d). 
The overlap was developed starting from [\plbfirst],
which in turn was motivated by [\kaplan] and [\frolovslavnov].
The formulation of vector-like theories which is the main
focus of this letter can be found in section 9 of [\npblong].
Work on 2d examples was successful [\twod]
and added confidence in the basic idea.
This induced a subsequent search for simplifications. One year
ago a compact action was found representing exactly massless
fermions on the lattice [\exactlymassless]. This compact action
is induced by integrating out all but one of the infinite 
number of fermions that are equivalent to the overlap. 
The fermions that are integrated
out are heavy and the ``miracle'' is the relative simplicity of the action
for the remaining massless fermion. That action confirms to a criterion
introduced many years ago by Ginsparg and Wilson (GW) [\GW] to assure
continuum-like Ward identities on the lattice. The compact action
produced by the overlap is the first known acceptable solution to the 
GW requirement. To date, there are no other explicit solutions.
Related work can be found in [\hasen,\luscher].  

In spite of its compact form, the overlap-Dirac operator isn't easy
to work with numerically and its practical usefulness in dimensions
higher than two has been unclear until now. One purpose of this
letter is to present a new procedure to simulate the overlap-Dirac
operator on the computer. This procedure holds the promise
to be practicable in 3 and 4 dimensional gauge theories. 

One of the central issues in gauge theories is the spontaneous breakdown
of global chiral symmetries. This issue could not be addressed in 
a clean way on the lattice until now, but the new development
makes this possible. 

In three dimensions there are interesting models that are
parity invariant and have analogues of four dimensional 
global chiral symmetries [\appel]. 
Intensive investigations have indicated
that spontaneous breakdown of these symmetries takes place
if the (even) number of flavors is small enough. 

In the continuum spontaneous chiral symmetry breakdown is correlated
with an enhancement in the spectral density of the Dirac operator
around zero. Clean checks of this were impossible on the lattice
before the new actions. 

In four dimensions it has been suspected that configurations
consisting of superposed instantons/anti-instantons are responsible
for this accumulation of almost zero eigenvalues. In two dimensions
it is quite likely that this is actually true [\casherneub]. In three
dimensions there are no instantons, but there is another source
of low lying levels. Level crossings at zero must occur in backgrounds
connecting topologically distinct pure gauge configurations.
These crossings also happen in four dimensions. Until now, the 
level crossings could not be realized on the lattice since the
needed symmetries were broken by the regularization. 

Using the new action and the new procedure I shall exhibit below a 
level crossing in three dimensional $SU(2)$ gauge theory
with two flavors. At the moment I see no major difficulties left
to overcome on the way to a full dynamical simulation of this
exactly ``chirally symmetric'' model. Generalizations to four
dimensions also no longer appear prohibitive. It is possible
that this new procedure will revolutionize the way fermions
are treated on the lattice. 

The compact overlap-Dirac operator introduced in [\exactlymassless] is
$$
D={{1+\gamma_5 \epsilon (H)}\over 2},\eqno{(1)}$$
where $H=\gamma_5 D_W$. There is some freedom
in choosing $D_W$. The simplest choice is to take 
$D_W$ as the Wilson-Dirac operator with hopping parameter $\kappa$ set at 
$\kappa={1\over{2d-2}}$. This is the choice adopted henceforth. 

The first attempt to use $D$ directly, rather than the original
overlap formula [\npblong], 
was made in two dimensions for a $U(1)$ gauge theory by
T-W Chiu [\twchiu]. He used a Newton iteration to find $\sqrt {H^2}$.
His method required storage of the full matrix $H$ in memory. It is
easy to bypass the computation of $\sqrt{H^2}$ and establish an
iteration for $\epsilon (H)$ directly:
$$
{1\over X_{k+1}} = {1\over 2} \left ( X_k +{1\over X_k }\right );~~
k=0,1,2,...;~X_0=H.\eqno{(2)}$$
Observe now that the iteration can be ``solved'' by the replacement
$W_k ={{1-X_k}\over{1+X_k}}$. I assume that $X_0$ has no 
eigenvalue equal to $0$ and if this is true it will be also true
of all $X_k$. Equation (2) immediately leads to $W_k =(W_0 )^{2^k}$.
Setting $n=2^{k-1}$, we write the solution to (2) as $X_k = f_n (H)$. 
The function $f_n (z)$ is given by 
$$
f_n (z) ={{(1+z)^{2n} -(1-z)^{2n} }\over { (1+z)^{2n} + (1-z)^{2n}}}.
\eqno{(3)}$$
$f_n (H)$ is a rational approximant to $\epsilon (H)$ of the
Pad{\' e} type. 
The following identity is easily derived by calculating all the poles
of $f_n (z)$ and their residues:
$$
f_n (z) ={z\over n} \sum_{s=1}^n {1\over{
z^2 \cos^2 {\pi\over {2n}} (s-{1\over 2}) +\sin^2 {\pi\over {2n}} 
(s-{1\over 2})}}.\eqno{(4)}$$
$f_n (H)$ is relatively easy to calculate using, for
example, a CG algorithm for the inversions. Equation (4) has been
previously derived by applied mathematicians [\higham] in a different
context.

The computational cost is not much
higher than a single inversion since the inversions for all $s$
are related by shifts [\multiplemass]. The convergence of the CG
iteration is controlled by the $s=1$ term. There is no reason
to require $n$ to be a power of 2 any more, and $f_n (H)$ is a
truncation of $\epsilon(H)$ for any $n\ge 1$. In practice, below, 
I used $n=48$ to get ten digits precision when
acting on random vectors. Essentially
all the precision required of the CG was maintained. Since $f_n$ is
odd all that one needs to check is how close 
${{\| f_n (H) b \|^2}\over {\| b \|^2}}$ 
is to unity for randomly chosen vectors $b$. Since $\epsilon(\lambda
H)=\epsilon(H)$ for positive $\lambda$ we also have 
scaled versions of $f_n$, with slightly
different convergence properties.
Since physics is concentrated in the spectrum of $H$ close to
zero it is not recommended to choose a $\lambda$ to ensure optimal
overall convergence. For simplicity, I have kept $\lambda=1$ in what
follows. 

The formula for $H$ appropriate for three dimensions was obtained
by dimensional reduction along direction 4 from four dimensions
[\odddim]. The boundary conditions obeyed by the fermions are picked
periodic in direction 4 and anti-periodic in directions 1,2,3. 
I used the real form of $H$ introduced in [\realsu2].
Starting from a single Dirac fermion in four dimensions we end up
with two Dirac fermions in three dimensions. Let us denote
$\gamma_5 \epsilon (H)$ by $V$. $V$ is orthogonal, and obeys
$\gamma_5 V \gamma_5 = \gamma_4 V \gamma_4= V^T$ and
consequently $[\gamma_4\gamma_5 , V]=0$. The first two
identities are of GW type. The three properties 
together constitute the 
lattice realization of the global ``chiral'' $SU(2)$ symmetry
known in the continuum. The
reality of $V$ implies that parity is conserved at the action
level. In the continuum, it is believed that 
parity does not break down spontaneously,
but the global $SU(2)$ symmetry does. 
The breaking is accompanied by two 
massless ``pions''. 

The crossing we wish to see occurs when we interpolate
smoothly unit link variables to a lattice pure gauge
configuration of special type. The gauge transformation
defining the latter is a 
discretized form of a known nontrivial continuum map
from the three torus in the continuum to the $SU(2)$ group manifold
[\su2anom]. The interpolation parameter is denoted by $t$
and goes from 0 to 1. There is a built in symmetry in the
configuration under $t\rightarrow 1-t$. Thus, the crossing
is expected exactly at $t=.5$. 

To see the crossing I compute the lowest few eigenvalues of
$1+{1\over 2}(V+V^T )$ by a CG based variational method [\ritz].
The lowest state is found to be doubly degenerate. It corresponds
to two conjugate eigenvalues of $V$, $e^{\pm i\theta}$. For
$t$ close to .5, $\theta$ is close to $\pi$, and the crossing
can be easily seen from a plot of the square root of 
the eigenvalue $1+\cos \theta (t)$
as a function of $t$. The flow is shown in the Figure 1. 

The crossing takes place on lattices as small as $6^3$ and has been
confirmed by the indirect method of [\su2anom]. That method is based
on the Herzberg -- Longuet-Higgins effect [\berry]. Without any effort
I increased the volume to $16^3$. At this point $H$ is a 
$32,768\times 32,768 $ real matrix. In QCD it has been claimed by
Lepage and co-workers that physical results can be obtained on
lattices as small as $6^4$ [\lepage]. Such a lattice has a 
complex fermionic matrix of size $15,552\times 15,552$. This
is comparable to our example. 

Many refinements of the procedure outlined here are possible.
Lurking behind the scenes are Chebyshev polynomials and they
have appeared in this context before,
in the analysis of another truncation of the overlap,
studied in [\trunca,\truncb]. The light fermions in the latter
truncation are usually referred to as domain wall
fermions. There are many reasons why I suspect that the method
presented here is superior to domain wall fermions.
Space does no permit a more thorough discussion at this time. 

In this letter a new and direct approach to treat the overlap action
${{1+\gamma_5 \epsilon (H)}\over 2}$ on potentially
realistic lattices in three and four dimensions has been defined.
The feasibility of a serious numerical study in $d=3$ of a certain class
of interesting models has been established. These results
open up many directions for future research.

\figure{1}{\captionone}{5.0}

{\bf Acknowledgments: } This work was supported in 
part by the DOE under 
grant \#DE-FG05-96ER40559.
I am grateful to Claudio Rebbi
for providing me with the 
qcdf90 package [\rebbi]
which I adapted to my needs in this work.
I am grateful to Artan Bori{\c c}i for sending me a PS
copy of reference [\higham] and for related comments.

\vfill\eject

\centerline{\bf References.}
\medskip 
\item{[\plbfirst]} R. Narayanan, H. Neuberger, 
Phys. Lett. B302 (1993) 62.
\item{[\kaplan]} D. B. Kaplan, Phys. Lett. B288 (1992) 342.
\item{[\frolovslavnov]} S. A. Frolov, A.A. Slavnov, 
Phys. Lett. B309 (1993) 344.
\item{[\npblong]} R. Narayanan, H. Neuberger, Nucl. Phys. B443 (1995) 305.
\item{[\twod]} R. Narayanan, H. Neuberger, P. Vranas, 
Phys. Lett. B353 (1995) 507.
\item{[\exactlymassless]} H. Neuberger, Phys. Lett. B417 (1998) 141.
\item{[\GW]} P. Ginsparg, K. Wilson, Phys. Rev. D25 (1982) 2649.
\item{[\hasen]} P. Hasenfratz, V. Laliena,
F. Niedermayer, hep-lat/9801021.
\item{[\luscher]} M. L{\" u}scher, hep-lat/9802011.
\item{[\appel]} T. Appelquist, D. Nash, Phys. Rev. Lett. 64 (1990) 721.
\item{[\casherneub]} A. Casher, H. Neuberger, Phys. Lett. B139 (1984) 67.
\item{[\twchiu]} Ting-Wai Chiu, hep-lat/9804016.
\item{[\higham]} I have become aware of this after posting the
original version of this letter. See equation (6.5) in the review by
Nicholas J. Higham, ``The Matrix Sign Decomposition and its Relation
to the Polar decomposition'', in Linear Algebra and Appl., 
Proceedings of ILAS Conference ``Pure and Applied Linear Algebra:
The New Generation'', Pensacola, March 1993 and references therein.
\item{[\multiplemass]} A. Frommer, S. G{\" u}sken, T. Lippert, B. N{\" o}ckel,
K. Schilling, Int. J. Mod. Phys. C6 (1995) 627.
\item{[\odddim]} Y. Kikukawa, H. Neuberger, Nucl. Phys. B513 (1998) 735.
\item{[\realsu2]} H. Neuberger, hep-lat/9803011.
\item{[\su2anom]} H. Neuberger, hep-lat/9805027.
\item{[\ritz]} B. Bunk, K. Jansen, M. 
L{\" u}scher, H. Simma, DESY-Report(Sept. 94);
T. Kalkreuter, H. Simma, Comp. Phys. Comm. 93 (1996) 33.
\item{[\berry]} G. Herzberg, H. C. Longuet-Higgins,
Disc. Farad. Soc. 35 (1963) 77.
\item{[\lepage]} G. P. Lepage, hep-lat/9607076.
\item{[\trunca]} H. Neuberger, Phys. Rev. D57 (1998) 5417.
\item{[\truncb]} Y. Kikukawa, H. Neuberger, A. Yamada, hep-lat/9712022.
\item{[\rebbi]} I. Dasgupta, A. R. Levi, V. Lubicz, 
C. Rebbi, Comp. Phys. Comm. 98 (1996) 365.

\vfill\eject 
\end